\documentstyle[aps,epsf,multicol,floats]{revtex}
\makeatletter
\def\thepage{\@arabic\c@page}
\def\@pnumwidth{2em}
\def\REVTeX{REV\TeX}
\makeatother

\begin{document}

\title{Free energies of crystalline solids: a lattice-switch Monte Carlo method}

\author{A.D. Bruce, N.B. Wilding \& G.J. Ackland}
\address{Department of Physics and Astronomy, The University of
Edinburgh\\
Edinburgh, EH9 3JZ, Scotland, United Kingdom} 
\maketitle

\makeatletter
\global\@specialpagefalse
\def\@oddhead{\REVTeX{} 3.0\hfill Released November 10, 1992}
\let\@evenhead\@oddhead
\def\@oddfoot{\reset@font\rm\hfill \thepage\hfill
\ifnum\c@page=1
  \llap{\protect\copyright{} 1992
  American Institute of Physics}%
\fi
} \let\@evenfoot\@oddfoot
\makeatother

\begin{abstract}

We present a method for the direct evaluation of the difference between the
free energies of two crystalline structures, of different symmetry. The method 
rests on a Monte Carlo procedure which allows one to sample along a path,
through  atomic-displacement-space,  leading from one structure to the other by way
of an intervening  transformation that switches one set of lattice vectors for another. The configurations of both structures
can thus be sampled within a single Monte Carlo process, and the difference
between their free energies evaluated directly from the ratio of the measured
probabilities of each. The method is used to determine the  difference between
the free energies of the {\it fcc} and {\it hcp} crystalline phases of a system of hard spheres.


\end{abstract}


\newpage
\twocolumn

One of the fundamental tasks of theoretical condensed matter physics is to 
understand the observed structures of crystalline materials in terms of 
microscopic models of the atomic interactions. The principles involved are
well known: one needs to evaluate which of the candidate structures has
the lowest free energy for given (model and thermodynamic) parameters. In
practice the task is rather less straightforward. Conventional Boltzmann
importance sampling Monte Carlo (MC) methods do not yield the free energy
\cite{binder}. It
is therefore customary to resort to integration methods (IM) which determine
free energies by integrating  free-energy {\em derivatives} measured at intervals along
a parameter-space-path connecting the system of interest to a reference
system whose free energy is already known. This procedure has been used widely,
and with ingenuity \cite{frenkelladd}. Nevertheless it leaves much to be desired.
In particular, to
determine the {\em difference} 
between the free
energies of two phases one has to relate {\em each} of them separately to
some reference system, with uncertainties which are not always transparent, and
which can be significant on the scale of the free energy difference of interest. Clearly,
one would prefer a method which focuses more directly on this
difference. The elements of such a strategy are to be found in the
extended-sampling techniques  pioneered by Torrie and Valleau \cite{torr;vall},
and recently revitalized in the multicanonical method of Berg and Neuhaus
\cite{bn}.
The key concept underlying this method is that of a 
configuration-space-path comprising  the macrostates of some chosen
macroscopic property ${\cal M}$. The method utilizes a sampling distribution
customized to even out the probabilities of different ${\cal M}$-macrostates.
In principle, it allows one to  sample along a
path (whose canonical probability is generally extremely small) chosen to connect 
the distinct
regions of configuration space associated with two phases; the difference between the free
energies of the two phases can then be obtained directly from the ratio of the
probabilities with which the system is found in each of the two regions. This
idea has been applied in the investigation of the phase behavior of
ferromagnets \cite{bn},  fluids \cite{nbw}, and lattice gauge theories
\cite{lgt}. However its application to {\em structural} phase behavior faces a
distinctive problem:  finding a path  that links the regions of configuration
space associated with two different crystal structures \cite{grsadb},   {\em
without} traversing regions of non-crystalline order, which present problems
\cite{blocks}
for even multicanonical Monte-Carlo studies. We show here how one may
construct such a path, and use it for direct high-precision measurement of
free-energy differences of crystal structures.

The idea is simple; we describe it first in general and qualitative terms.
The atomic position coordinates are  written, in the
traditions of lattice dynamics, as the sum of a lattice vector \cite{useoflattice}, and a
displacement vector. The configurations associated with a particular structure are
explored by  MC sampling of the displacements. Given any configuration
of one structure one may identify a  configuration 
of the other, by {\em switching} one  set  of lattice vectors for the
other, while keeping the displacement vectors {\em fixed}. Such lattice
switches can be incorporated into the MC procedure by regarding the
lattice type as a stochastic variable. 
Lattice switches have an intrinsically
low acceptance probability, since typically they entail a large energy cost.
But the multicanonical method can be  used
to draw the system 
along a path comprising the  macrostates of this `energy cost',
and thence into a region of displacement-space in which
the `energy cost' is low, and the lattice switch can be implemented. The net result is a MC
procedure which visits both structures in the course of a single simulation,
while never moving out of the space of crystalline configurations. 
The method is, we believe, potentially
very general. We illustrate it here by using it to determine the difference
between the free energies of the two close-packed structures ({\it fcc} and
{\it hcp}) of a system of hard spheres.  This problem has a long history
\cite{aldercarteryoung}. The difference between the free energies
(effectively, the {\em entropies}) 
is extremely small, and 

recent IM 
studies have disagreed
on its value \cite{woodcock,bolfrenk}.
It thus provides an exacting
and topical testing ground
for  the method proposed here \cite{widersignificance}.

We consider a system of $N$ particles with spatial coordinates $\{\vec{r}\}$. 
The particles are confined in a fixed volume $V$, with periodic
boundary conditions \cite{novolumechange}. We make the decomposition
\begin{equation}
\vec{r}_i=\vec{R}_i+\vec{u}_i
\label{eq:decomposition}
\end{equation} 
where the vectors $\vec{R}_i, i=1\ldots N
\equiv\{\vec{R}\}_\alpha$ define the sites of a lattice of type
$\alpha$ (here, either {\it fcc} or {\it hcp}). 
Clearly there are many transformations that will map one set of vectors into the
other; the mapping we have chosen is explained in Fig. 1(a),(b): it exploits the fact
that the two structures differ only in the stacking pattern of the close-packed
planes.

We define a  partition function  (and free energy) 
associated with the structure $\alpha$ by \cite{generalnotation}
\begin{eqnarray}
Z(N,V,T,\alpha)& =& 
\int_{\{\vec{u}\} \in \alpha} \prod_i [  d\vec u _i ]
\exp\left[- \Phi(\{\vec{u}\},\alpha) \right] \nonumber \\
&\equiv& \exp\left[- F_{\alpha} (N,V,T)/kT\right]
\label{eq:canpartdefa}
\end{eqnarray} 
where $\Phi$ represents the  dimensionless
configurational energy. In the present context
\begin{equation}
\Phi(\{\vec{u}\},\alpha) \equiv \Phi(\{\vec{r}\})= \left\{
\begin{array}{ll}
0 & \mid \vec{r}_i -\vec{r}_j\mid >\sigma\,\, \forall i,j \\
\infty & \mbox{otherwise}
\end{array}
\right .
\label{eq:Phidef}
\end{equation} 
where $\sigma$ is the hard sphere diameter. The $\alpha$-label attached to the
integral in
Eq.~(\ref{eq:canpartdefa}) signifies that it 
must include only contributions from configurations within the subspace
associated with the structure $\alpha$\cite{associated}.

Consider now the canonical ensemble with  probability distribution

\begin{equation}
P(\{\vec{u}\},\alpha\mid N,V,T) = \frac
{\exp\left[- \Phi(\{\vec{u}\},\alpha) \right]}
{Z(N,V,T)} 
\label{eq:rawcanconfdist}
\end{equation} 
where $Z(N,V,T) \equiv\sum_{\alpha}Z(N,V,T,\alpha)$. The probability that the
system will be found to have structure $\alpha$ provides a  measure of the
associated partition function:
\begin{eqnarray}
P(\alpha\mid N,V,T)& \equiv&
\int_{\{\vec{u}\} \in \alpha} \prod_i [  d\vec u _i ]
P(\{\vec{u}\},\alpha\mid N,V,T) \nonumber\\
&=& \frac{Z(N,V,T,\alpha)}{Z(N,V,T)}
\label{eq:phaseprobs}
\end{eqnarray} 

The difference between the free-energies of the two structures may be
thus be expressed as
\begin{equation}
F_{\mbox{\it hcp}}(N,V,T)-F_{\mbox{\it fcc}}(N,V,T) \equiv NkT \Delta f =
kT \ln{{\cal R}}
\label{eq:deltaf}
\end{equation} 
where, 
\begin{equation}
{\cal R} = \frac{Z(N,V,T,\mbox{\it fcc})} {Z(N,V,T,\mbox{\it hcp})} 
= \frac{P(\mbox{\it fcc}\mid N,V,T)} {P(\mbox{\it hcp}\mid N,V,T)}
\label{eq:Rdef}
\end{equation} 
This identification is useful {\em only} if one can
devise a MC procedure that will actually visit the configurations
$\{\vec{u}\},\alpha$ with the probabilities prescribed by Eq.~(\ref{eq:rawcanconfdist}).
To do so one must deal with the ergodic
block against lattice switches (`updates' of the lattice label, $\alpha$):
almost 
invariably such a switch maps an accessible  configuration of 
one structure  onto an inaccessible configuration of the 
other (one which violates the hard-sphere constraint
implied by Eq.~(\ref{eq:Phidef})). Fig. 1(b) provides an example. The 
resolution is to {\em bias} the sampling procedure  so as to favor 
the occurrence of configurations which transform {\em without}
violating this constraint. To do so we define an {\em overlap order parameter}

\begin{equation}
{\cal M}(\{\vec{u}\}) 
\equiv 
M(\{\vec{u}\},{\mbox{\it hcp}}) 
-M(\{\vec{u}\},{\mbox{\it fcc}}) 
\label{eq:overlapdef}
\end{equation} 
where $M(\{\vec{u}\},\alpha)$ counts the number of pairs
of overlapping spheres associated with the configuration
$\{\vec{u}\},\alpha$
(again, see Fig. 1(b)). 
Since  $M(\{\vec{u}\},\alpha)$ 
will necessarily be zero for any set of displacements $\{\vec{u}\}$ {\em actually visited} when the
system has lattice $\alpha$, the order parameter ${\cal M}$
is necessarily $\ge 0$ ($\le 0$) for realizable configurations of the {\it fcc}
({\it hcp}) structure. The  displacement configurations with ${\cal M}=0$
are accessible in {\em both} structures and thus offer no
barrier against lattice switches. Accordingly the set of ${\cal
M}$-macrostates provides us with the required `path' connecting the two phases,
through a lattice-switch at ${\cal M}=0$.
To pick out this path we must sample from the biased configuration distribution
\begin{equation}
P(\{\vec{u}\},\alpha \mid N,V,T, \{\eta \})\propto
P(\{\vec{u}\},\alpha \mid N,V,T)e^{ \eta ({\cal M}(\{\vec{u}\}) )} 
\label{eq:multicanconfdist}
\end{equation} 
where $\{\eta\} \equiv \eta ({\cal M}), {\cal M} =0,\pm 1,\pm 2 \ldots$ define a
set of multicanonical weights \cite{bn}, which have to be determined such that
configurations of all relevant 
${\cal M}$-values 
are sampled. Once this is done, one can measure the weighted distribution
of ${\cal M}$-values, and reweight (unfold the bias) to determine
the true canonical form of this distribution:
\begin{equation}
P({\cal M}\mid N,V,T) \propto
P({\cal M} \mid N,V,T, \{\eta\})e^{-\eta ({\cal M})}
\label{eq:candist}
\end{equation} 
Finally, the difference between the free energies of the two structures may be
read off from this distribution through the identification (cf Eqs.~(\ref{eq:deltaf}),(\ref{eq:Rdef}))
$\Delta f = N^{-1} \ln {\cal R}$, with
\begin{equation}
{\cal R}= \frac{
\sum_{{\cal M} > 0} P({\cal M}\mid N,V,T)
}{
\sum_{
{\cal M} < 0} P({\cal M}\mid N,V,T)
}
\label{eq:Rvalue}
\end{equation} 

We have implemented this procedure to study systems of N=216, 1728 and 5832
hard spheres (forming, respectively 6, 12 or 18 close-packed layers). The
volume  $V$ was chosen such that the fraction of space filled, $\rho$,
satisfies $\rho/\rho_{cp} = 0.7778$ \cite{whythisdensity}, where $\rho_{cp}\equiv 0.7404$ is the
space filling fraction in the closest packing limit. The MC procedure 
entails sampling the displacement variables $\{\vec{u}\}$ and the lattice label
$\alpha$. The variables $\{\vec{u}\}$ were updated by drawing new values from a
top-hat distribution \cite{useoftophat} 
and accepting them provided they
satisfy the hard sphere constraint; the lattice switches 
were attempted (and accepted with probability $1/2$) only when the system is in
the ${\cal M}=0$ macrostate.
The weights (which enable the system to reach this
special macrostate) were obtained using methods explained elsewhere
\cite{grsadb}. We allowed typically $2\times10^4$ sweeps for equilibration and
up to $5\times10^7$ sweeps for final sampling runs on the largest system size. 
The simulations were conducted on DEC ALPHA workstations using  overall
some 800 hours CPU time.

Fig. 2 shows the measured overlap distribution for the
system of $N=1728$ spheres; the inset shows the probability on a logarithmic
scale, exposing the enormity of
the entropic `barrier' (probability `trough') that the
multicanonical weighting enables us to negotiate.
The difference between the free energies of the two structures is identifiable
immediately and transparently from the ratio of the integrated weights  of the two
essentially gaussian peaks. Our results for this system and other system sizes
are gathered together in Table 1, along with the results of other authors.

From Table 1 it is apparent that the present work greatly refines the 
largely inconclusive results of the original 
IM study \cite{frenkelladd}. Our results are consistent with
--though substantially more precise than-- very recent IM studies of 
Bolhuis and Frenkel \cite{bolfrenk}. They are inconsistent with the result reported by Woodcock
\cite{woodcock}, given that $\Delta f$
is believed to {\em decrease} as the density is reduced, towards melting
\cite{pressurecalc}.
While we have not attempted an explicit analysis of the finite-size behavior,
the close agreement between our
results for $N=1728$ and $N=5832$ indicates that the latter should provide an extremely good
estimate of the thermodynamic limit, confirming the stability of the {\it fcc}
structure at this density.

Our principal concern here, however, is with the {\em general} lessons that can
be learned  about the method introduced in this work. The precision we have achieved
with this method is self-evidently a significant advance on that of IM studies.
Admittedly, this level of precision has entailed substantial
processing {\em time}, principally because of the relative slowness of the
diffusive exploration of the multicanonically weighted configuration space.
But the point is  that the procedure is {\em practicable} \cite{notaeons}, with
a computational strategy that is, we suggest, less complex and more transparent
than that  of IM. Thus, for example, the method described in \cite{frenkelladd}
involves integration (of a mean-square displacement) along a parameter-space-path connecting
each structure to a reference system, comprising an Einstein-model of the same structure; the MC integral then
has to be combined with the known free energy of the Einstein model, a virial
correction, and a correction to the virial correction, before taking the
difference between the results for the two structures. The uncertainties in all
the contributions have to be assessed separately. By contrast, the present
method focuses directly on the quantity of interest (the relative weights of the
peaks in Fig. 2); and the precision with which
it is prescribed is defined by standard MC sampling theory.  


Finally we comment briefly on the more general applicability of the method. 
For systems other than hard spheres, the role of the overlap order parameter is
played by the {\em energy barrier} encountered in the lattice switch; the
generalization of the  weighting procedure should be straightforward. It seems
unlikely that many problems will require the level of precision needed here,
where the two phases are so finely balanced. However some circumstances will
not generally prove as favorable. In the present case one can readily identify
a lattice-to-lattice mapping which {\em guarantees} no overlaps
(high-energy-cost interactions) amongst subsets of the atoms (those lying
within the {\em same} close-packed plane). The optimal form of mapping may not
always be so evident. It may also prove advantageous to relax
the constraint imposed in Eq.~(\ref{eq:decomposition}) that the coordinates
$\{\vec{u}\}$ represent {\em identical} displacement patterns in the two structures. 
At the cost of only  little extra computational complexity the true
displacements may be represented as structure-dependent (even site-dependent)
{\em functions} of a (still common) set of coordinates  $\{\vec{u}\}$. It
should then be possible to  ensure that a typical displacement pattern in one
structure maps onto a typical pattern in the other. 
These matters are the
subject of ongoing study.

 \newpage
\newcounter{abc}
\renewcommand{\thefigure}{\arabic{figure}\alph{abc}}

 \renewcommand{\thefigure}{\arabic{figure}\alph{abc}}
 \setcounter{abc}{1}
 \begin{figure}[h] 
   \leavevmode
   \epsfxsize=70mm 
   \begin{center}
     \vspace*{-1cm}
     \epsffile{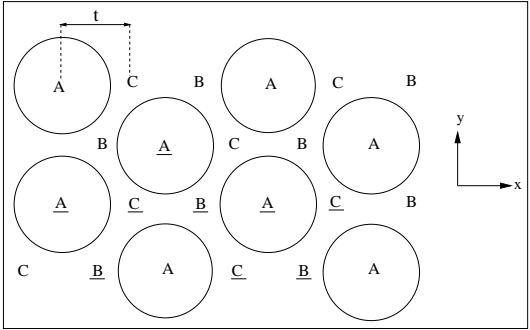}
    \end{center}
   \caption
   { 
Schematic representation of the close-packed structures. 
The points marked A show the positions of the sites in one close-packed ($xy$) layer;
the circles show the boundaries of spheres
occupying these sites in an ideal (zero-displacement) structure. The points
marked B and C show the projections of sites in other layers (stacked along
the $z$-axis) onto the $xy$
plane;  the {\it fcc}  and {\it hcp} structures entail sequences of type $ABCA 
\ldots$ and  $ABAB  \ldots$  respectively. The {\em lattice switch}  from {\em
fcc} to {\em hcp} entails {\em translations} of the close-packed planes, 
as detailed in Fig. 1(b).
   }
   \label{fig:figonea}
 \end{figure}
 \addtocounter{abc}{1}
 \addtocounter{figure}{-1}



\begin{figure}[t] 
  \leavevmode
  \epsfxsize=70mm 
  \begin{center}
    \vspace*{-0.2cm}
    \epsffile{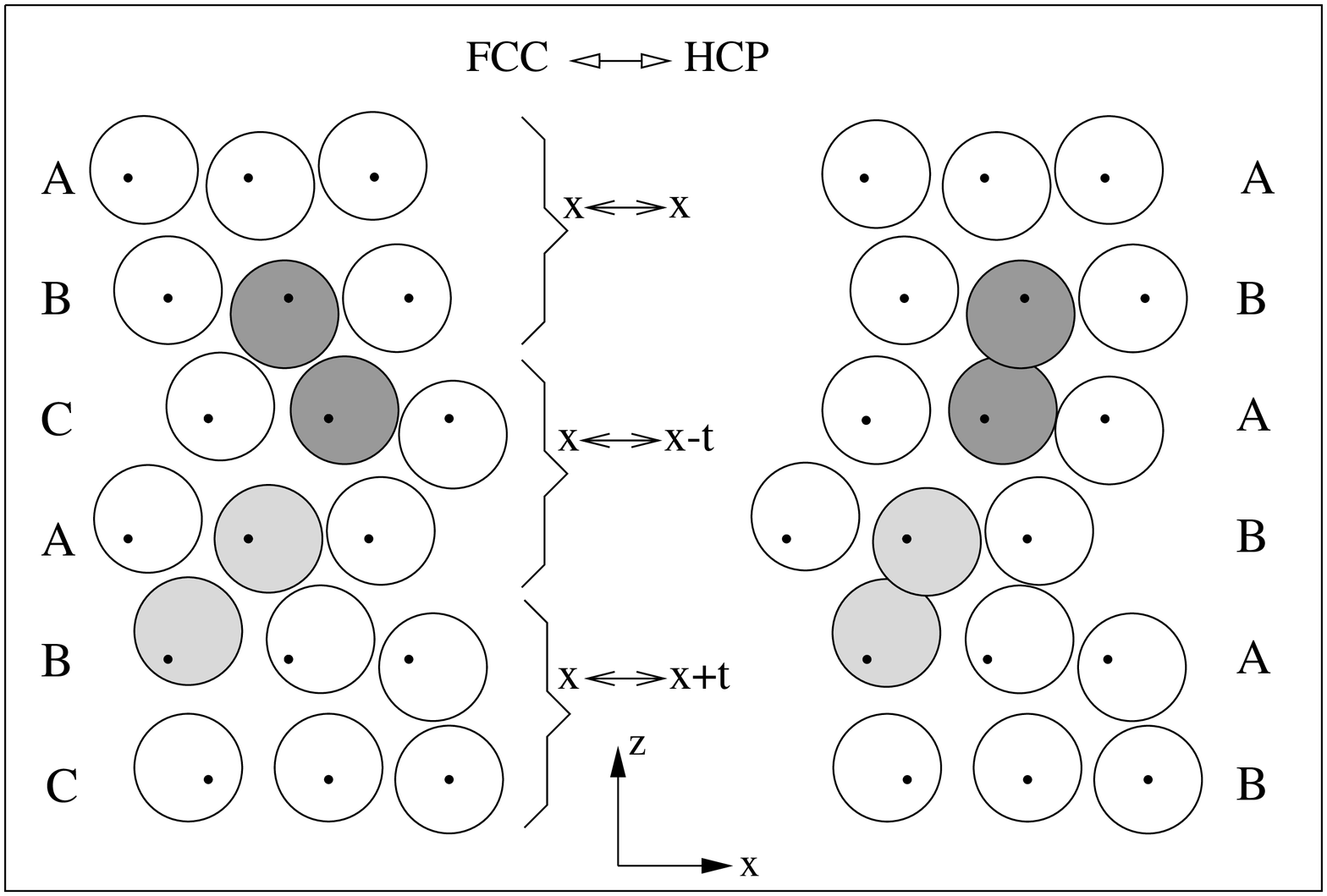}
  \end{center}
  \caption
  { 
[Left side] The positions of the spheres in an arbitrary 
configuration of the {\it fcc} structure, projected onto the $xz$ plane.
We show 6 layers, with 3 spheres in each; the sites in the top 3 layers (A,B,C)
correspond to those marked (and underlined) A,B,C in Fig. 1(a). \protect\newline
[Center] The action of the lattice switch: reading from the top, the
first two layers (A,B) of the {\em fcc} structure are invariant; the next two  (C,A)
layers are translated along the $x$ direction by $-t$; and the final two (B,C)
layers are translated by $+t$, where $t$ is identified in Fig. 1(a). \protect\newline
[Right side] Projections of
the spheres in the resulting {\it hcp} arrangement. Here, the
displacements $\{\vec{u}\}$, realizable in the {\it fcc} structure,
give two overlapping pairs of spheres (shaded)
in the {\it hcp} structure so that 
${\cal M}(\{\vec{u}\})= 2$ in this case.
Note that
the picture is {\em schematic}: in particular, the  density shown here is much
lower than that chosen for the present study.
  }
  \label{fig:figoneb}
\end{figure}
\setcounter{abc}{0}


\begin{figure}[t]
  \leavevmode
  \epsfxsize=80mm 
  \begin{center}
    \vspace*{-0.5cm}
    \epsffile{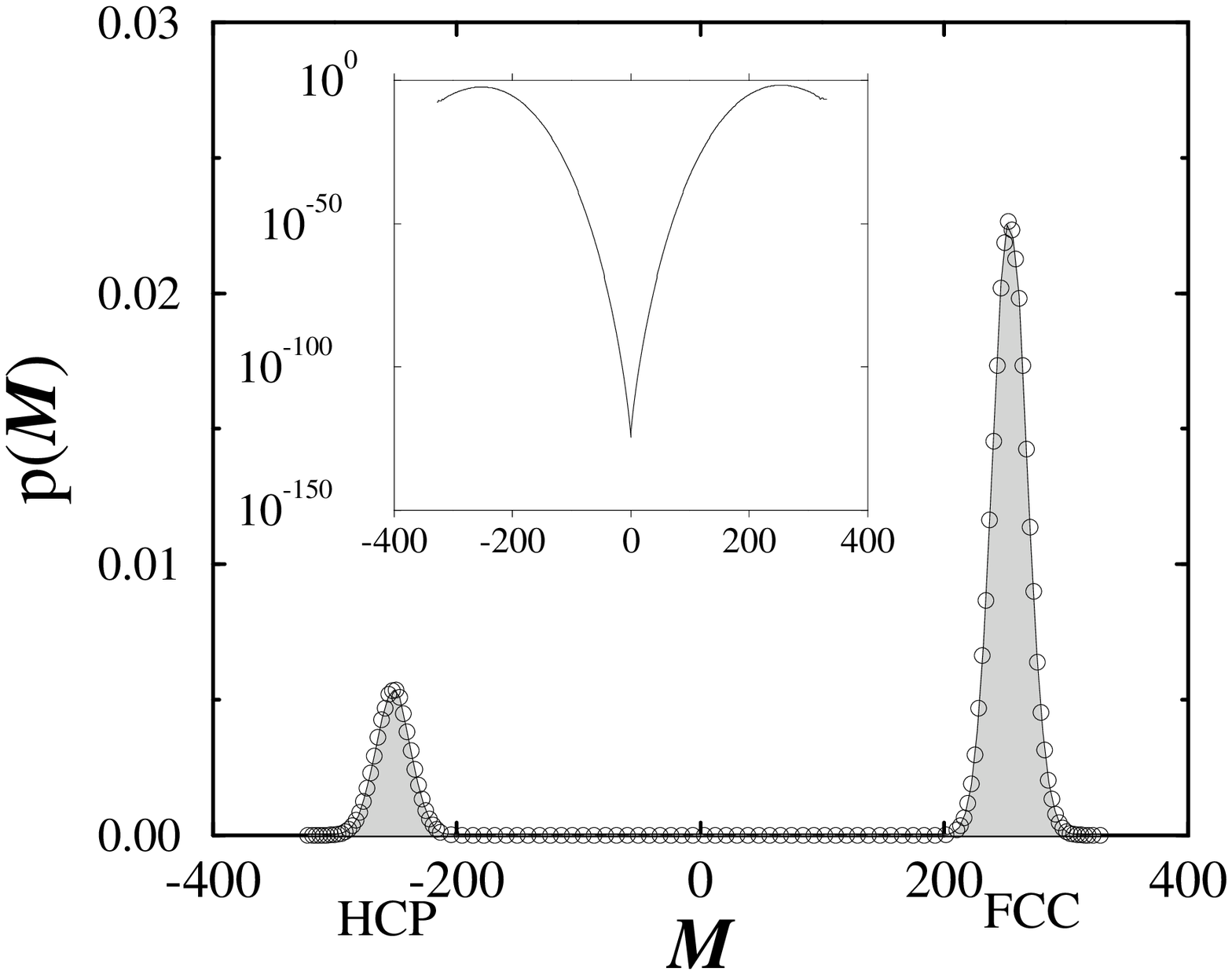}
  \end{center}
  \caption
  { 
The distribution of the overlap parameter ${\cal M}$ for a system of 
N=1728 spheres; the inset shows the distribution
on a logarithmic scale. The statistical uncertainties are smaller than
the symbol size.
The free-energy difference $\Delta f$
is identified from the logarithm of the ratio ${\cal R}$  of the weights
of the two peaks (Eq.(\ref{eq:Rvalue})). The smallness of $\Delta f$ allows both peaks to be displayed
on one linear scale in this case.
  }
  \label{fig:figtwo}
\end{figure}

\narrowtext
\begin{table}[t]
  \begin{center}
    \vspace*{-0.3cm}
    \begin{tabular}{|lrrrl|}
$\rho/\rho_{cp}$&N  & \multicolumn{2}{c}{$\Delta f \times 10^5$ } & Ref.  \\ 
\tableline
0.7360 &216&           90 &(135) & \cite{frenkelladd}     \\     
0.7360 &12000&        500&(100)    & \cite{woodcock}           \\     
0.7360 &12906 &        90&(20)& \cite{bolfrenk}  \\     
0.7778 &1152   &       -120&(180) & \cite{frenkelladd}     \\     
0.7778 &216     &      101& (4)  & PW            \\
0.7778 &1728     &     83& (3)  & PW            \\
0.7778 &5832      &    86&(3)  & PW
    \end{tabular}
    \label{tab:coexden}
  \end{center}
  \caption
  {
Results for the difference between the free energy of {\it hcp} and {\it fcc}
structures, as defined in Eq.~(\ref{eq:deltaf}) with associated uncertainties
in parentheses. Results attributed to Ref.~\protect\cite{frenkelladd} were deduced by
combining the separate results for {\it fcc} and {\it hcp} given there. 
PW signifies the present work. The PW
error bounds were computed from the statistical uncertainties in the weights of
the peaks in $P({\cal M})$.
  }
\end{table}


\begin{references}
\bibitem{binder} K. Binder, {\it J. Comp. Phys.} {\bf 59} 1 (1985)
\bibitem{frenkelladd} D. Frenkel \& A.J.C. Ladd, {\it J. Chem. Phys} {\bf 81}, 3188 (1984).
\bibitem{torr;vall} G.M. Torrie  \& J.P. Valleau,   {\it Chem. Phys. Lett.} {\bf 28}, 578 (1974).
\bibitem{bn}  B.A. Berg \& T. Neuhaus,  {\it Phys. Lett. B} {\bf 267}, 249 (1991); {\it Phys. Rev. Lett.} {\bf 68}, 9 (1992).
\bibitem{nbw} N.B. Wilding,   {\it Phys. Rev. E} {\bf 52},602 (1995).
\bibitem{lgt} B. Grossmann, M.L. Laursen, T. Trappenberg \& U.J. Wiese, {\it Phys. Lett. B} {\bf  293}, 175 (1992).
\bibitem{grsadb} G.R. Smith \& A.D. Bruce, {\it Phys. Rev. E} {\bf 53}, 6530
(1996) apply multicanonical methods to a structural phase transition which
involves {\em no change of symmetry} where an appropriate path is 
identified simply by ${\cal M} =\rho$.
\bibitem{blocks} Eg: the ergodic block associated with re-crystallization.
\bibitem{useoflattice} We use the term `lattice vector' a little loosely: we
mean the set of vectors identified by the orthodox crystallographic lattice, convolved with the
orthodox basis.
\bibitem{aldercarteryoung} B.J.Alder, B.P. Carter \& D.A. Young, {\it Phys Rev} {\bf 183}, 831 (1969).
\bibitem{woodcock} L.V. Woodcock, {\it Nature} {\bf 384}, 141 (1997).
\bibitem{bolfrenk} P.G.Bolhuis \& D. Frenkel (unpublished).
\bibitem{widersignificance} The hard sphere system has a wider significance.
Y.Choi {\it et al},{\it J. Chem. Phys.} {\bf 99} 9917 (1993) 
show that predictions for the phase diagram of a Lennard Jones solid
depend extremely sensitively on the hard-sphere free-energy difference $\Delta
f$ computed here. Colloids  provide
experimental realizations of 
near-hard-sphere systems: 

P.N. Pusey {\it et al}, {\it J. Phys: Condens. Matter} {\bf 6} A29 (1994).
\bibitem{novolumechange} In common with previous studies we
work at constant {\em volume}, and constrain
the $c/a$ ratio in the {\it hcp} structure to its 
closest-packing value. Generalization to the constant {\em pressure}
ensemble is straightforward in principle. 
\bibitem{generalnotation} We use a general notation; formally the properties
 of the hard-sphere system are independent of $T$.
\bibitem{associated} In the MC context the configurations {\em
associated} with a given structure are  identified as the
set which are actually {\em accessed} in a simulation 
initialized within the set. 
\bibitem{whythisdensity} This value of $\rho$ was chosen to coincide with one of
those studied in Ref~\cite{frenkelladd}.
\bibitem{useoftophat} This choice of sampling procedure ensures that the center
of mass is effectively fixed. For consistency
the width of the top-hat distribution must
be large compared to the range of displacements actually {\em
accepted}.
\bibitem{pressurecalc} This is the implication of studies of the pressure in the
two structures:  
B.J.Alder,  D.A. Young, M.R. Mansigh \& Z.W. Salsburg, {\it J. Comp. Phys} {\bf 7}, 361 (1971).
\bibitem{notaeons} The time required is measured on a scale of hours rather than the
eons required  if one were to attempt such a `direct' method {\em without} the
multicanonical strategy provided here: recall the scale on the inset of Fig. 2.




\end{references}
\end{document}